\documentclass{article}

\usepackage{arxiv}

\usepackage[utf8]{inputenc} 
\usepackage[T1]{fontenc}    
\usepackage{hyperref}       
\usepackage{url}            
\usepackage{booktabs}       
\usepackage{amsfonts}       
\usepackage{nicefrac}       
\usepackage{microtype}      
\usepackage{lipsum}
\usepackage{float}
\usepackage{graphicx}
\usepackage{amsmath}

\title{Image Processing Tools for Financial Time Series Classification}

\author{
  Bairui Du\\
  Department of Computer Science\\
  Faculty of Engineering\\
  University College London\\
  Gower Street, London, WC1E 6BT, UK.\\
  \texttt{bairui.du.18@ucl.ac.uk} \\
  \And
  Delmiro Fernandez‑Reyes\\
  Department of Computer Science\\
  Faculty of Engineering\\
  University College London\\
  Gower Street, London, WC1E 6BT, UK.\\
  \texttt{delmiro.fernandez-reyes@ucl.ac.uk} \\
  \And
  Paolo Barucca\thanks{Corresponding Author: Paolo Barucca, Financial Computing and Analytics Group, Department of Computer Science, University College London, Gower Street, London, WC1E 6BT, UK. Email: p.barucca@ucl.ac.uk}\\
  Department of Computer Science\\
  Faculty of Engineering\\
  University College London\\
  Gower Street, London, WC1E 6BT, UK.\\
  \texttt{p.barucca@ucl.ac.uk}
}

\begin{document}
\maketitle

\begin{abstract}
The application of deep learning to time series forecasting is one of the major challenges in present machine learning. 
We propose a novel methodology that combines machine learning and image processing methods to define and predict market states with intraday financial data. 
A wavelet transform is applied to the log-return of stock prices for both image extraction and denoising.
A convolutional neural network then extracts patterns from denoised wavelet images to classify daily time series, i.e. a market state is associated with the binary prediction of the daily close price movement based on the wavelet image constructed from the price changes in the first hours of the day. 
This method overcomes the low signal-to-noise ratio problem in financial time series and get a competitive prediction accuracy of the market states ‘Up’ and ’Down’ of financial data as tested on the S\&P 500. 
\end{abstract}


\section{Introduction}
\label{sec:Introduction}
Time series prediction is a challenge for many complex systems, yet in finance predictions are hindered by the very nature of how financial markets work.  
In efficient markets, the opportunities for stock price predictions leading to profitable trades are supposed to rapidly disappear. 
In the growing industry of high-frequency trading, the competition over extracting predictions on stock prices from the increasing amount of available information for performing profitable trades is becoming more and more severe. 
With the development of big data analysis and advanced deep learning methodologies, traders hope to fruitfully analyse market information, e.g. price time series, through machine learning. 
Spot prices of stocks provide a simple snapshot representation of a financial market. 
Stock prices fluctuate over time, affected by numerous factors, and the prediction of their changes is at the core of both long-term and short-term financial investing. 
The collective patterns of price movements are generally referred to as market states. 
As a paramount example, when stock prices follow an upward trend, it is called a bull market, and when stock prices follow a downward trend is called a bear market \cite{procacci2019forecasting}. 

However, both in bullish and bearish market trends, there are lots of noisy oscillation, requiring analysts to apply noise reduction methodologies to extract meaningful predictions over trends. 
The objective of this study is to test a general time-series prediction model which extracts a denoised wavelet image from time series in order to leverage over convolutional neural network (CNN) architectures. 
We apply continuous wavelet transforms to the log return of financial time series and convert it to a greyscale wavelet transform spectrum. 
Then we build one shallow and one deep convolutional neural network (CNN) model and train them with spectrum image as input to capture hidden patterns.
The main novelty and contribution of our study is to define stock market states based on intraday financial time series whilst also providing accurate predictions, demonstrating the ability of image processing tools to overcome the low signal-to-noise ratio of financial time series, and providing a promising toolbox for the analysis of noisy time series found in many complex systems.

\section{Background and Related Work}
\label{sec:Background}
Market states forecasting is based on the analysis of historical data \cite{selvin2017stock}, yet as summarised in \cite{murphy1999technical}, the quest for predictions in financial data needs to take into account a series of empirical laws of financial markets: "Market action discounts everything; Prices move in trends; History tends to repeat itself". 
In this paper, we look at different financial indicators and analyse them to identify patterns, trends, periods or cycles.
Compared to other time series, financial time series display a significant amount of uncertainty and unpredictability \cite{tsay2010financial}. 
As the raw price time series will often contain a trend, using log-returns instead of prices is an established method to transform raw data \cite{giles2001noisy,procacci2019forecasting,siripurapu2014convolutional}, returns being a good scale-free summary of the outcome of investment decisions withing a given time interval. 
In most quantitative financial research and applications, log-returns are regarded as more tractable, as having more robust and characteristic statistical properties, e.g. probability distribution over a given period, and, from a practical point of view, it is possible to quickly produce multi-periods returns from single period ones \cite{procacci2019forecasting,tsay2010financial}. 

Financial forecasting can be framed as a signal processing problem \cite{giles2001noisy} on which neural networks can be applied to provide testable solutions. 
In \cite{giles2001noisy} the authors consider log-returns of stock prices, denoise the log-return time series, and apply self-organizing map (SOM) - an unsupervised learning method to learn the distribution of a set of patterns without any class information - to make predictions.

More recently, researchers have tried to apply deep learning methods such as convolutional and recurrent neural networks to predict the behavior of stock markets.
Advanced machine learning methods have a high representation power for empirical asset pricing, being able to re-create arbitrarily complex non-linear multi-variate functions and not requiring an arbitrary feature selection pre-processing that could dilute the information content of the original time series \cite{gu2018empirical}. 
In \cite{gu2018empirical} the authors perform a comparative analysis of different methods such as simple linear, penalised linear principal components regression (PCR), partial least squares (PLS), regression trees and random forests \cite{gu2018empirical}, and find that regression trees have the best prediction accuracy \cite{gu2018empirical}. 
They consider a continuous variable regression problem rather than a classification problem. 
At odds with results in image and bio-metric pattern recognition where the deeper the neural network the better, they find that shallow learning outperforms deeper learning, the reasons for this phenomenon being (1) the low signal-to-noise ratio of financial data and (2) the comparative scarcity of the data for the price prediction problem \cite{gu2018empirical,siripurapu2014convolutional}. 
They also proved that, compared to the traditional prediction methods, machine learning provides an improved description of the behavior of expected returns.

In order to reduce the complexity and improve the accuracy of the forecast, we consider the stock forecasting problem not as a regression problem, but as a classification problem for determining a market state \cite{procacci2019forecasting}. 
In \cite{procacci2019forecasting} the authors give a model-based clustering method which clusters financial time series via a maximum-likelihood model. 
Their clustering procedure uses a likelihood measure adjusted for temporal coherence. 
This procedure is shown to be numerically efficient and suitable for high dimensional datasets, alternating (a) the update of the network structure constructed by the TMFG-LoGo algorithm and (b) the assignment of points to clusters in a time-consistent manner through dynamic programming, i.e. using the Viterbi algorithm \cite{procacci2019forecasting}. The model both identifies the current market state, i.e. bull or bear market, and yields predictions for future market states. 
The final values of accuracy for these two cases are above 50\%.

Our study combines the time series prediction task explored in deep learning applications in finance and the market state identification problem investigated within statistical approaches to derive a predictive classification of the intraday behavior of financial indices, introducing a methodology for time series analysis applicable in the broader context of complex systems.

\section{Methodology}
\label{sec:Methodology}
\begin{figure*}[htbp]
    \centering
    \includegraphics[width=\textwidth]{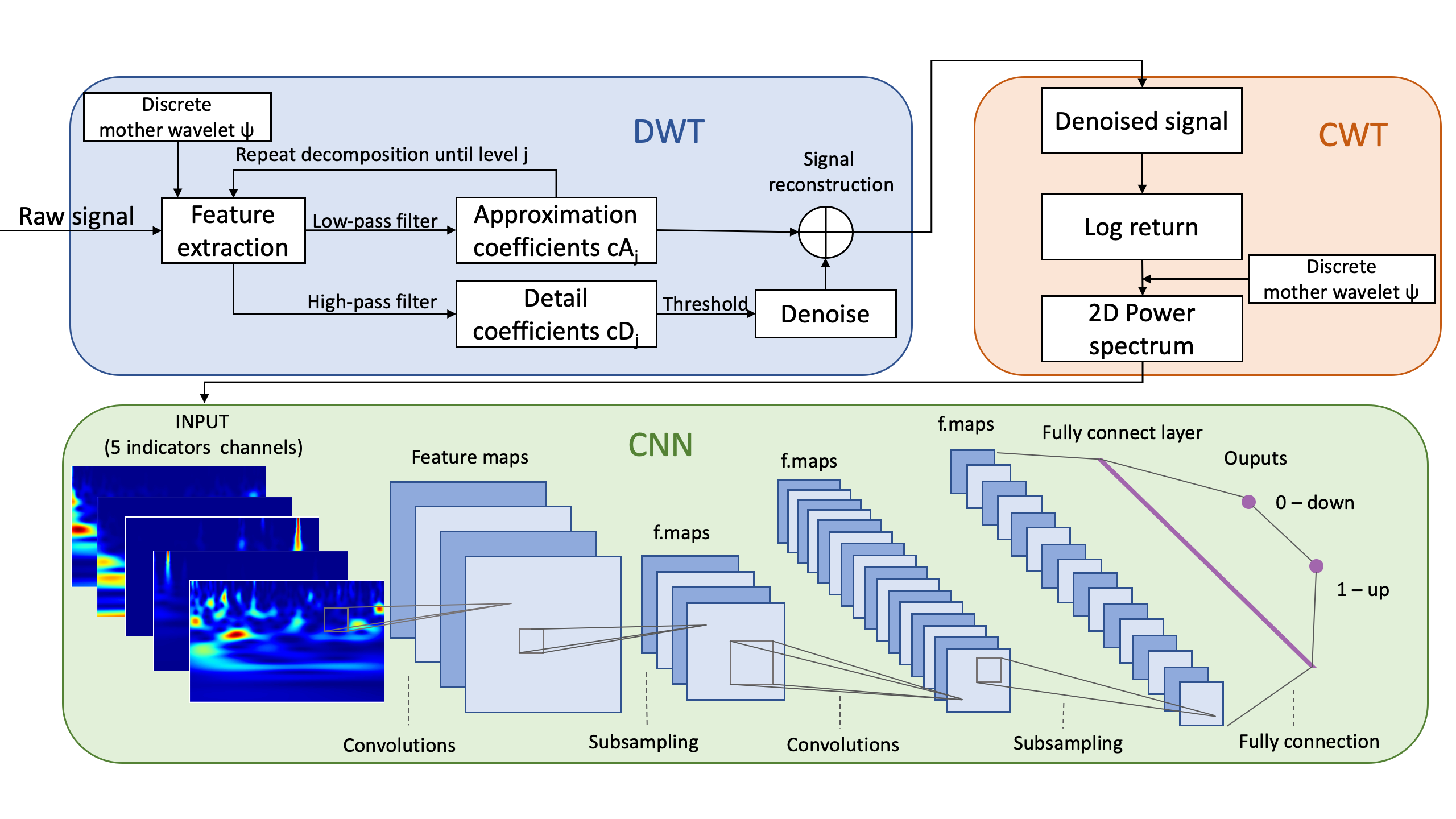}
    \caption{Flowchart of using wavelet transform and CNN to do financial time series prediction}
    \label{fig:whole process}
\end{figure*}
\subsection{Wavelet Transform}
Fourier transform is a powerful data analysis tool that represents any complex signal as a sum of sines and cosines and transforms the signal from the time domain to the frequency domain \cite{portnoff1980time}. 
Nevertheless,  Fourier transform can show which frequencies are present in the signal, whilst it cannot show when these frequencies appear. 
The short-time Fourier transform divides original signal into several parts using a sliding window to fix this problem\cite{griffin1984signal}. 

Wavelet transform is a more suitable method for analysing dynamic signals, as it identifies the existing frequencies in the original signal and also when these frequencies appear and disappear by controlling the scale change of the wavelet. Therefore, wavelet transform yields an high resolution in both the frequency domain and time domain. 
A wavelet is a rapidly decaying, wave-like oscillation that has zero mean and is localised in both time and frequency space\cite{farge1992wavelet}. 
Unlike sines waves, which extend and repeat to infinity, a wavelet exists for a finite duration. 

Real world signals do not always change slowly, and they often oscillate or expose transient changes. 
In financial time series analysis, these abrupt changes can be associated with turning points which can be crucial for stock market forecasting. 
However, high frequency noise may hinder the detection of these turning points. 
In most cases, noise is modelled as Gaussian white noise \cite{diebold1998elements, procacci2019forecasting}.
Fourier transform does not represent these abrupt changes efficiently, because it does not consider time-dependence in the signal decomposition.
Daubechies points out that the wavelet transform can be used to analyse time series that contain non-stationary power at many different frequencies\cite{daubechies1990wavelet}. 

\subsection{Discrete Wavelet Transform} 
\begin{figure}[htbp]
\centering
    \includegraphics[width=10cm]{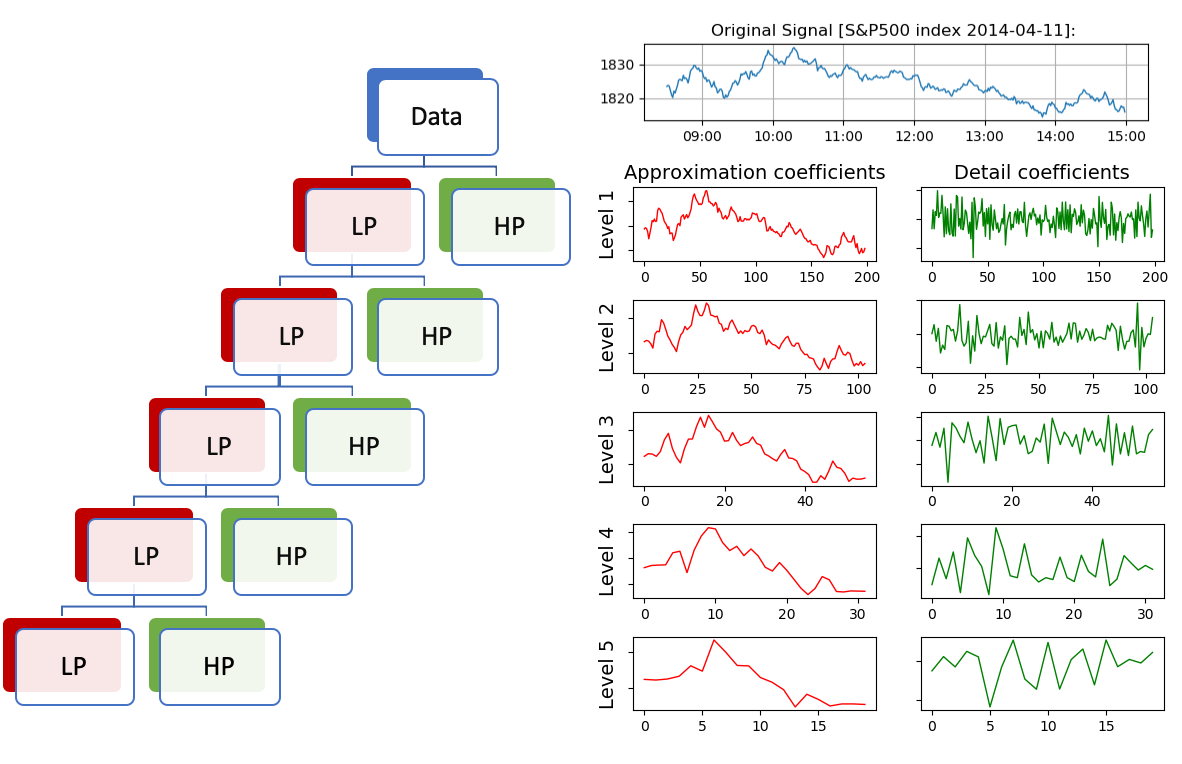}
    \caption{DWT decomposition into approximation coefficients and details coefficients (wavelet cofficients)}
    \label{fig:DWT decomposition}
\end{figure}

The wavelet transform threshold denoising method was first proposed in \cite{donoho1994ideal}. 
The basic idea of wavelet threshold denoising is that after the signal is transformed (e.g. using Mallat algorithm) \cite{shensa1992discrete,donoho1995adapting}, it is further decomposed into approximate coefficients and detail coefficients. 
The detail coefficients are also called wavelet coefficients. The wavelet coefficients with larger amplitudes are assumed to be significant for representing the original signal, while coefficients with smaller amplitudes are generally associated with noise\cite{shensa1992discrete}.
The threshold denoising method finds a suitable threshold, retains the wavelet coefficients larger than the threshold, filters the wavelet coefficients smaller than the threshold accordingly, and then restores the denoised signal according to the processed wavelet coefficients \cite{donoho1994ideal}.
 
Wavelet denoising can also be regarded as a low-pass filter. 
It removes high-frequency noise while retaining the characteristics of the low-frequency components of the signal. 
Hence, wavelet denoising is a combination of feature extraction and low-pass filtering. 
Wavelet transform has good time-frequency localization characteristics, which can preserve relevant signal spikes and sudden changes\cite{daubechies1990wavelet,tufekci2000feature}. 
Therefore, the wavelet transform is suitable for removing transient signals, as well as suppressing the interference of high-frequency noise, and effectively distinguishing low-frequency information from high-frequency noise.
\\
\textbf{Decomposition}\\

\begin{equation}
\begin{aligned}
x_{j+1, L}[n] &=\sum_{k=0}^{K-1} x_{j,L}[2 n-k] g[k] \\
x_{j+1, H}[n] &=\sum_{k=0}^{K-1} x_{j,L}[2 n-k] h[k]
\end{aligned}
\end{equation}

$x[n]$ is Discrete input signal, length N.
$g[n]$ is a low pass filter can filter out the high frequency part of the input signal and output the low frequency part.
$h[n]$ is High pass filter can filters out the low frequency part and outputs the high frequency part.
The theoretical maximum decomposition level for walevet transforms is $j= \left\lfloor\log _{2} n\right\rfloor$, where $n$ is the signal length. 
The larger the decomposition level, the more obvious the different characteristics of noise and signal performance, and the more conducive to the separation of signal and noise, yet for reconstruction, the higher the number of decomposition levels, the greater the reconstruction error. 
The available maximum decomposition level is related to the signal-to-noise ratio (SNR) of the original signal, but the SNR cannot be obtained from the measured data. 
In order to avoid the loss of signal distortion and achieve the best noise reduction effect, verbatim noise verification is performed on the DWT detail coefficients. 
The Daubechies wavelets, based on the work of Ingrid Daubechies \cite{daubechies1988orthonormal}, are a family of orthogonal wavelets defining a discrete wavelet transform and characterized by a maximal number of vanishing moments for some given support\cite{daubechies1990wavelet}. We use the db4 (Daubechies wavelet of order 4) wavelet in Fig \ref{fig:Four Wavelets} (c) as mother wavelet to do level 5 DWT. 
The approximation and detail coefficients are shown in Figure\ref{fig:DWT decomposition}. 
The approximation coefficients represent the output of the low pass filter (averaging filter) of the DWT. 
The detail coefficients represent the output of the high pass filter (difference filter) of the DWT. 
The difference between CWT and DWT is that DWT uses discrete values for the scale a and translation factor b. 
The DWT is only discrete in the scale and translation domain, not in the time-domain. 
On the left, we can see a schematic representation of DWT apply on the signal as a low pass filter at each level. 
The detail coefficients of Level 1 and 2 are more in line with white Gaussian noise characteristics.\\

\textbf{Threshold Denosing and Reconstruction}
\begin{figure}[htbp]
    \centering
    \includegraphics[width=\textwidth]{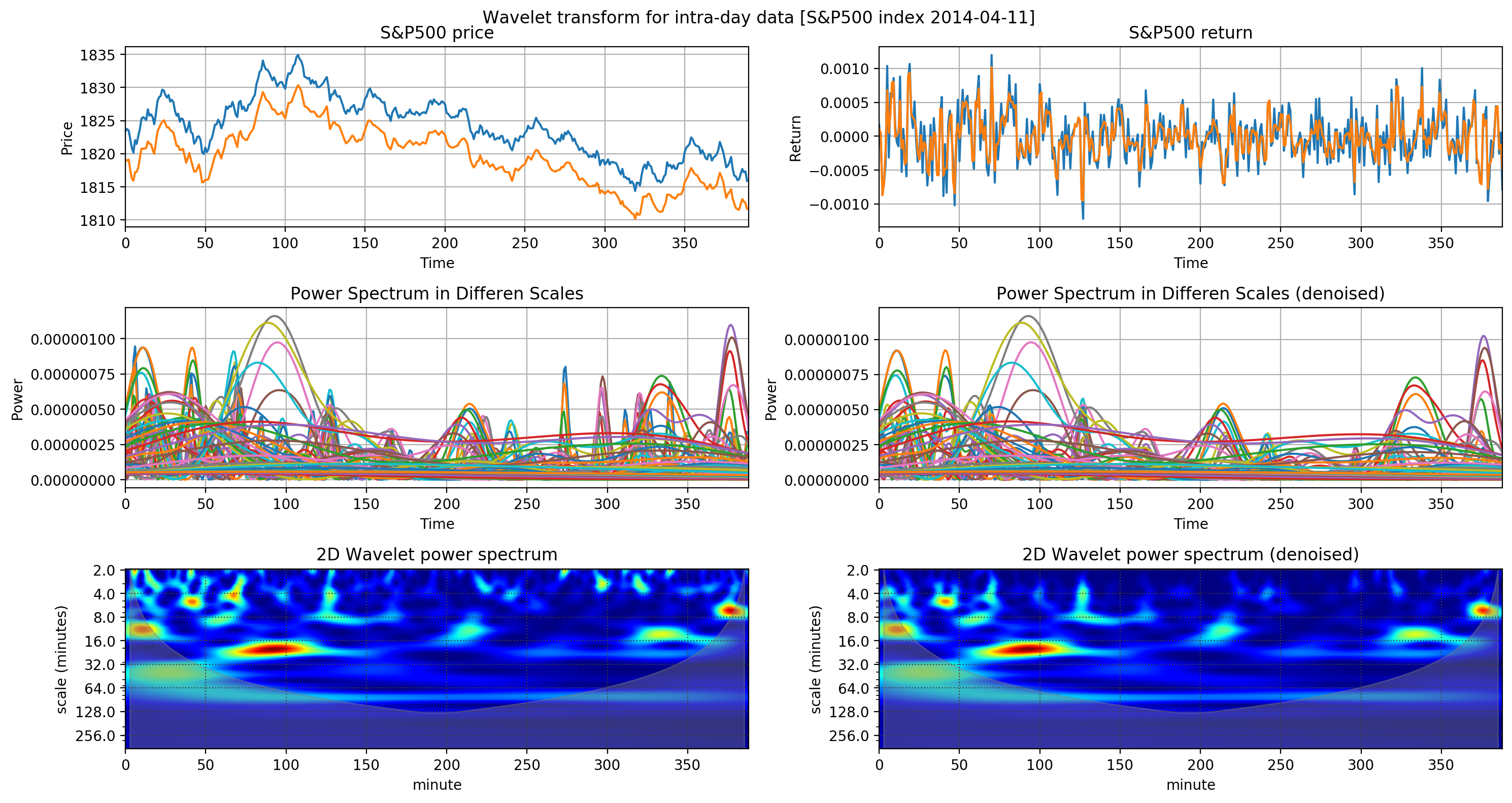}
    \caption{2D Wavelet Transform Spectrum with COI (Cone of Influence) after Denoising}
    \label{fig:S&P Wavelet}
\end{figure}

Threshold denoising performs nonlinear thresholding on the wavelet transform coefficients of the measured signal. 
The high-frequency coefficients of each layer from the $1$st to the $N$th layer are filtered by the threshold function, and the low-frequency coefficients of each layer are left unchanged.
The hard threshold \ref{eq:hard thresholding}is discontinuous at the threshold $\lambda$, which causes the denoised signal to oscillate around it, yet a hard threshold function can perform better than a soft threshold function in terms of mean square error. 
In this study we apply soft thresholding, prioritising the overall continuity of the wavelet coefficients ensured by a continuous function around the threshold \ref{eq:soft thresholding}. 

In the threshold processing function, the selection of the threshold $\lambda$ directly affects the effect of denoising. 
There are four types of thresholds for the wavelet transform threshold denoising method: general threshold rules, minimum maximum variance threshold, Stein’s Unbiased Risk Estimate (SURE) Rules, and heuristic threshold rules. 
In this study we use the rigrsure threshold method, which is an adaptive threshold selection based on the principle of Stein's unbiased likelihood estimation (quadratic equation). It first estimates the likelihood of different $\lambda$ values, and then minimizes it to get the selected threshold. 
 
Then, we reconstruct the signal from the filtered wavelet coefficients. 
The signal is reconstructed based on the low-frequency coefficients of the Nth layer of wavelet decomposition and the processed high-frequency coefficients of the first $N-1$ layers, so as to obtain new denoised values of the original signal.
The blue signals in Figure \ref{fig:DWT decomposition} Figure \ref{fig:S&P Wavelet} and are the original signal, and the orange signal is the signal after noise reduction. 
Finally, we perform the log-return transformation on the price signal. In Figure \ref{fig:S&P Wavelet} 

\subsection{Continuous Wavelet Transform}

Different applications may require different mother wavelets and there are two important wavelet transform concepts to be considered: scaling and shifting. Given a signal $\Psi\left(t\right)$, scaling refers to the process of stretching or shrinking the signal in time \cite{torrence1998practical}, which can be expressed in the following equation, 
\begin{equation}
\Psi\left(\frac{t}{s}\right){s>0}
\label{Eq:Scaling}
\end{equation}
${s}$ is the scaling factor that represents how much the signal is rescaled in time. The scale factor is inversely proportional to frequency. In a wavelet, there is a reciprocal relationship between scale and frequency with a constant of proportionality (COP) \cite{singh2017statistical}. The mother wavelet has a characteristic frequency band. Mathematically, the equivalent frequency is defined as follows, 
\begin{equation}
F_{e q}=\frac{C_{f}}{s \delta t}
\label{eq:COP}
\end{equation}
where $C_f$ represents the center frequency, $s$ the wavelet scale, and $\delta t$ the sampling interval. 

Continuous and Discrete Wavelet Transforms are two major wavelet analysis methods. CWT is mainly used in time-frequency analysis, and filtering Of time localized frequency components. DWT is ideal for denoising and compressing signals and images, as it helps represent many naturally occurring signals and images with fewer coefficients\cite{torrence1998practical}. The difference between CWT and DWT is how they discretize the scale and the translation parameters \cite{antonini1992image}. The CWT of a discrete sequence $x_n$ is defined as the convolution of $X$ with a scaled and translated version of 
wavelet $\psi_0(\eta)$\cite{torrence1998practical}:
\begin{equation}
    W_{n}(s)=\sum_{n^{\prime}=0}^{N-1} x_{n^{\prime}} \psi *\left[\frac{\left(n^{\prime}-n\right) \delta t}{s}\right]
    \label{CWT}
\end{equation}

Wavelets in Figure \ref{fig:Four Wavelets} are some of the well-known mother wavelets, and they have different sizes and shapes.   We use Morlet wavelet to generate power spectrum of the denoised signals. Equation \ref{eq:Morlet wavelet} can express the Morlet wavelet used in this thesis. Morlet is a plane wave modulated by a Gaussian where the $\omega_{0}$ is non-dimensional frequency, and $t$ is a  non-dimensional "time" parameter.

\begin{equation}
\Psi_{0}(t)=\pi^{-1 / 4} e^{i \omega_{0} t} e^{-t^{2} / 2}
\label{eq:Morlet wavelet}
\end{equation}

The output of CWT are coefficients, which are a function of scale, frequency and time\cite{antonini1992image}. The Higher the number of scales per octave the finer the scale discretization. When we do CWT, each scaled wavelet is shifted in time and compared with the original. Then repeating this process for all the scales results in coefficients that are a function of the scales and shift distance of wavelet \cite{torrence1998practical}. For example, a signal with 10000 samples analyzed with 50 scales will generate in 500,000 coefficients. In this way, oscillatory behaviour in signals can be characterised in more detail. 

\subsection{Feature Engineering}

\subsubsection{Data cleaning}
\begin{figure}[H]
    \centering
    \includegraphics[width=\textwidth]{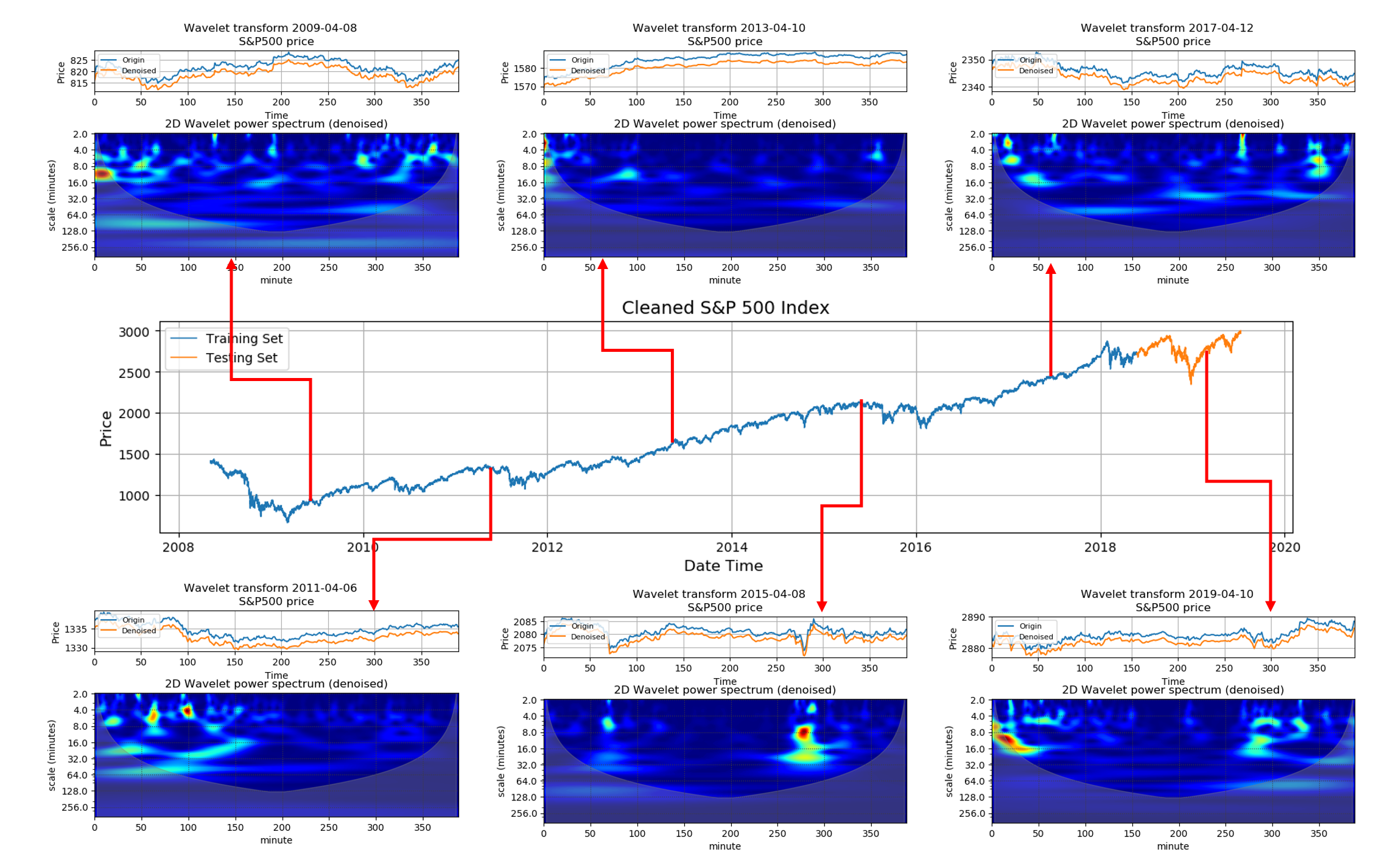}
    \caption{Wavelet spectrum in timeline}
    \label{fig:Spectrum in timeline}
\end{figure}

We consider 505 stocks in the tickers list of S\&P 500, widely regarded as the best single gauge of large-cap U.S. equities\cite{indices2019s}. \footnote{Eight of the stocks cannot be downloaded for reasons related to corporate reorganisation and name changing} First, we calculate the adjusted closing price and then we clean the data. The raw data of S\&P 500 index have 'Open', 'Close', 'High', 'Low' prices and 'Volume'. Given that only one-year out of ten includes 'Volume' data, we could not include it as a feature in our analysis. Some days, when the U.S. market closes early or delays opening, only have half-day data points. These points and weekends and holidays have been cleaned and deleted. Further, on a full trading day, we should have 390 data points, but 127 intraday data turn out to be incomplete. At this point, we have to deal with invalid values and missing values, which are due to the absence of transaction data for some minutes. The easiest way to do this is to replace the invalid and missing values with the sample mean, median, or mode of a variable. This method is simple but does not adequately consider the information already in the data, and the error may be significant. Sequential data in finance are significantly time-dependent. Therefore, for days when the intraday data is missing less than 20 data points, we fill NA/NaN values using the forward filling methods that propagate last valid observation forward to next valid. And if one-day data is missing more than 20 data points, we decide to simply not consider the intraday data from that day. This ensures that the input variables are consistent in time. The cleaned closing price is shown in Figure \ref{fig:Spectrum in timeline}, the blue part of ten years price data is used as the training data, and the red part of 1-year data is used as the test set.

Then we consider log-returns by taking the difference of log-transformed prices at two time consequent points in time from the raw price time series \cite{procacci2019forecasting}\cite{giles2001noisy}\cite{siripurapu2014convolutional}. 
Figure \ref{fig:S&P Wavelet} (a,b) show the raw price time series and log return time series.

\subsubsection{Wavelet power spectrum}
Figure \ref{fig:S&P Wavelet} shows the wavelet power spectrum. 
The abscissa displays time (390 minutes) and the y-axis here is log-scaled cause of the wide range of power spectrum values. 
The shaded region in the image is the cone of influence (COI). 
The scalogram is potentially affected by edge-effect artifacts and the unshaded region is a confidence area that should not be influenced by edge effects \cite{grinsted2004application,torrence1998practical}.
Wavelet transform is time-sensitive and provides an image representation from which the convolutional neural network can learn to recognise and extract hidden patterns regarding the underlying market state.

\subsubsection{Prediction target (y label)}
\begin{figure}[htbp]
    \centering
    \includegraphics[width = 7cm]{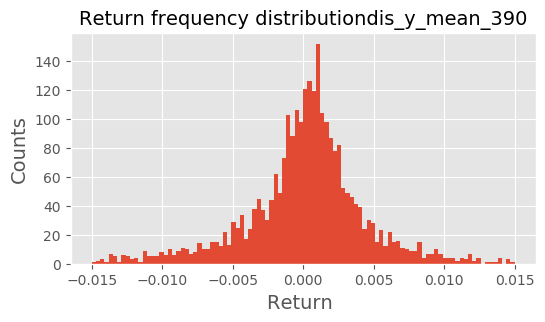}
    \caption{Return frequency distribution histogram of $y_mean_390$. X-axis is the return value and Y-axis is counts}
    \label{fig:Return_frequency_distribution}
\end{figure}

We have generated different y labels and shown in the Figure \ref{fig: other labels}and Figure \ref{fig:Return_frequency_distribution}. Figure \ref{fig:Return_frequency_distribution} is the log return between the average price from 1 to 360 minutes in the windows and the price at the last minute of the day. 
The reason we choose $y_{mean_390}$ as the classification label, is  that compared with other labels, it yields a broader distribution, which should translate into a greater margin for a good prediction from the convolutional neural network. 
Moreover, this label is more practical as, compared with the forecast of $y_{360_361}$, this label has more potential investment value.

\subsection{Feature selection}\label{Feature selection}

Feature selection refers to the identification of a set of prominent features for the task under study, selected according to a-priori criteria and preliminary investigations.
We used the Maximum Information Coefficient (MIC) method to select the top five training indicators as the selected input features. 
The mutual information (MI) of two random variables is a measure of the mutual dependence between the two variables\cite{asgarian2018impact}.
The formula for calculating the maximum information coefficient is as follows:
\begin{equation}
\operatorname{MIC}[x ; y]=\max _{|X||\mathbf{Y}<B|}\left(\frac{I[X ; Y]}{\log _{2}(\min (|X|,|Y|)}\right)
\end{equation}
Where X and Y represent two variables, respectively, and I[X; Y] represents mutual information for X and Y. Calculate the maximum information coefficient between features and categories to get the top 5 indicators in Table \ref{tab:MIC}. Because these prices are similar in minutes data, we only apply the MIC on closing other than all price. 
\begin{table}[H]
\centering
    \scalebox{0.8}{
        \begin{tabular}{|l|l|}
            \hline
            Indicator & Correlation coefficient \\
            \hline
            $index_{close}$ & 0.7590629756859861\\
            $EMA$    & 0.7440683555159783 \\
            $RSI$  & 0.73354493115159783 \\
            $MA60$  & 0.7241727805657679 \\
            $CORREL$& 0.712734946044241 \\
            \hline
        \end{tabular}}\\
        \caption{The top five indicator selected by the MIC method (PPirceChangeRatio) }
        \label{tab:MIC}
    \end{table}

\subsection{Experiment Design}
In this study, the market price predicting problem is treated as a classification problem with two classes, so that the output of the model is simply given by two labels, 'Up' or 'Down', that provide a prediction for the price movement during the last interval of a trading day. 
The trading time of a US stock market on a normal trading day is 390 minutes.
The 30 minutes after the opening and before the close are the most dramatic and uncertain 30 minutes\cite{bouchaud2002statistical}. 
In the half-hour before the close, traders may need to close their positions, make sure they execute an order, process new information from the day and act more or less rationally based on the daily trend, making predictions over the last price movements very challenging. 
In our experiment we use data from the first 360 minutes in a trading day as input to predict the closing market states, jumping beyond the unpredictable 30 minutes before the closing time. 
The label to 'Up' and 'Down' is calculated by comparing the average price at 360 minutes and the closing price of the stock market.
\begin{equation}
label_{y} = \left\{\begin{array}{l}
1,  (\sum_{i=1}^{360}\Psi(i))/360 < \Psi(390) \\\\
0,  (\sum_{i=1}^{360}\Psi(i))/360 \geq \Psi(390)
\end{array}\right.\\
\label{eq:y label}
\end{equation}
Where $\Psi$ is the price signal.

\subsubsection{Properties of CNN}
The reasons we choose CNN over other neural networks for image classification are three specific properties. 
The first property is locality, some patterns are much smaller than the whole image, and a set of neurons does not have to see the whole image to discover the pattern.
The second property is parameter sharing, as same patterns may appear in different regions and these patterns may have the same shape and also yield the same parameters, i.e. network weights and biases. In CNNs neurons can share parameters to reduce their overall number. 
Finally, image recognition subsampling. We can subsample the pixels reducing the number of parameters needed to process the image.
\section{Result}
\begin{figure}[htbp]
    \centering
    \includegraphics[width = 6cm]{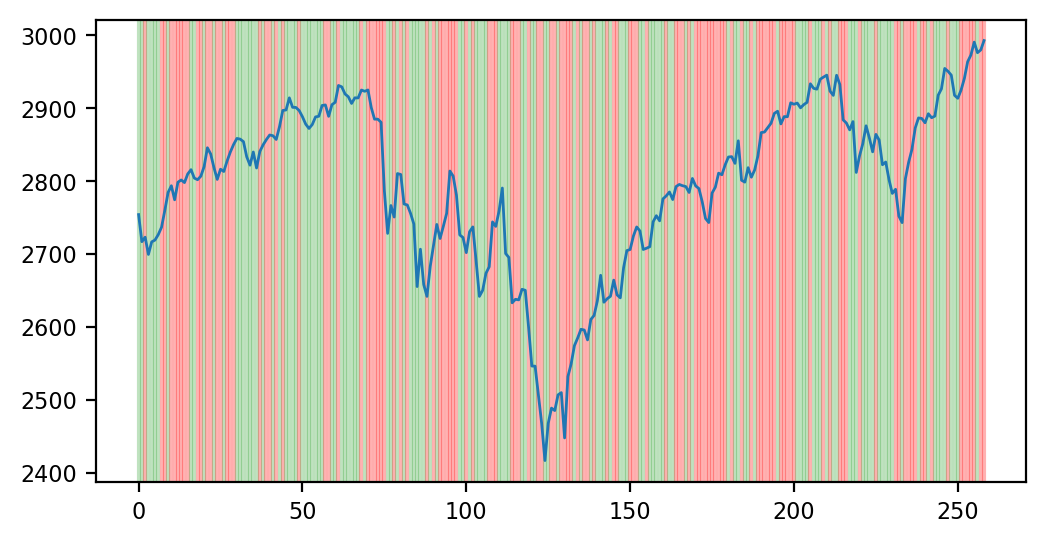}
    \includegraphics[width = 6cm]{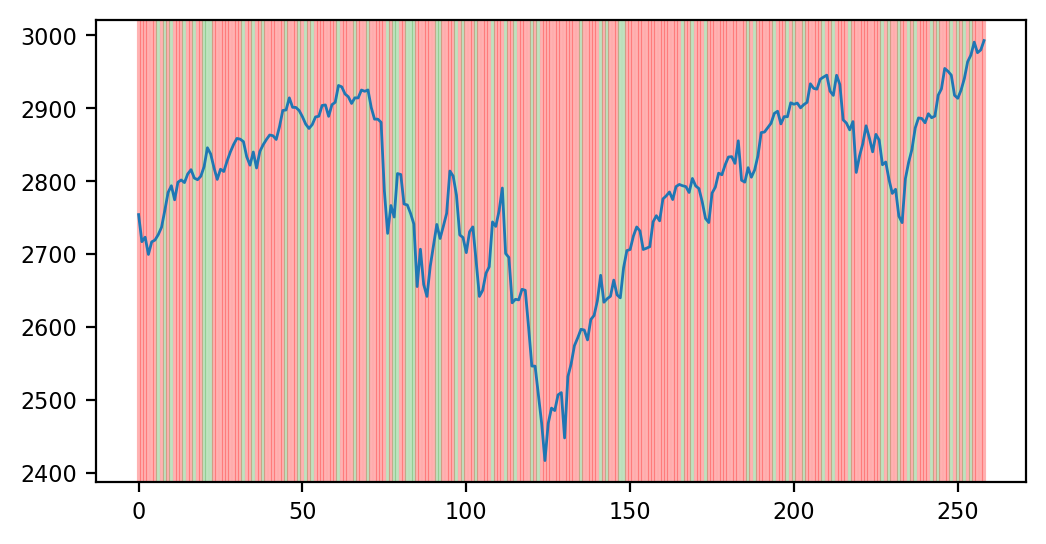}
    \caption{The prediction result of our model (the left picture) and random prediction (the right picture). Green corresponds to a correct prediction, and Red to a wrong prediction.}
    \label{fig:prediction result}
\end{figure}

From the Table\ref{Tab:confusion_matrix} \ref{Tab:confusion_matrix_random} below, we can see that this algorithm has a high accuracy rate compared to random prediction.
Figure \ref{fig:prediction result} more clearly shows the prediction results of our method and random prediction on the test set. The green corresponds to correct predictions, and the red corresponds to wrong predictions.

Although the true negative rate (TNR) of the denoised signal decreases, compared with the original signal, the overall prediction accuracy and F1 score improves. 
Taking into account the noise of financial data and the variability of samples, different experimental designs and label choices will affect the accuracy. 
The performance of this model on the S\&P500 index is competitive. 
\begin{table}[H]
    \centering
    \scalebox{0.8}{  
    \begin{tabular}{|lll|lll|}
    \hline
    \hline
    \multicolumn{3}{|c|}{Denoised signal}     & \multicolumn{3}{|c|}{Raw signal}    \\ \hline
                   & Actual 1      & Actual 0     &                 & Actual 1      & Actual 0  \\
    Predicted 1    & TP = 104      & FP = 67      & Predicted 1     & TP = 70       & FP = 50   \\
    Predicted 0    & FN = 43       & TN = 45      & Predicted 0     & FN = 77       & TN = 62   \\ \hline
    Loss           & \multicolumn{2}{l|}{0.722588} & Loss            & \multicolumn{2}{l|}{0.893730}\\
    Accuracy       & \multicolumn{2}{l|}{0.577220} & Accuracy        & \multicolumn{2}{l|}{0.507772}\\
    TPR            & \multicolumn{2}{l|}{0.707483} & TPR             & \multicolumn{2}{l|}{0.476190}\\
    TNR            & \multicolumn{2}{l|}{0.401786} & TNR             & \multicolumn{2}{l|}{0.553571}\\
    F1 score       & \multicolumn{2}{l|}{0.654088} & F1 score        & \multicolumn{2}{l|}{0.524345}\\ \hline
    \end{tabular}
    }

    \caption{Confusion matrix and accuracy.}
    \label{Tab:confusion_matrix}
    \end{table}
\begin{table}[H]
\centering
    \scalebox{0.8}{  
    \begin{tabular}{|lll|}
    \hline
    \hline
    \multicolumn{3}{|c|}{Random prediction} \\ \hline
                      & Actual 1       & Actual 0      \\
    Predicted 1       & TP = 58        & FP = 50       \\
    Predicted 0       & FN = 89        & TN = 62       \\ \hline
    Loss              & \multicolumn{2}{l|}{/}          \\
    Accuracy          & \multicolumn{2}{l|}{0.463320}   \\
    TPR               & \multicolumn{2}{l|}{0.394558}   \\
    TNR               & \multicolumn{2}{l|}{0.553571}   \\
    F1 score          & \multicolumn{2}{l|}{0.454902}   \\ \hline
    \end{tabular}
     }

    \caption{Random prediction}
    \label{Tab:confusion_matrix_random}
\end{table}

\section{Discussion}
The methodology developed in this study classifies the stock market on a given day into two basic market states, 'Up' and 'Down', providing better-than random predictions for the market state, measured as the price movement in the last time interval of a given trading day.
This study defines and addresses a specific classification task where machine learning can achieve superhuman performances in finance, i.e. the prediction of the final price movement in a given market day based on the return time series observed earlier in the day, denoised and wavelet transformed in order to be processed as an image by a convolutional neural network. 

The model uses discrete wavelet transform to reduce noise. Then a continuous wavelet transform is applied to the denoised signal to generate a spectrogram. We performed the above processing for multiple indicators to obtain multiple spectrograms, and then convert them into a multi-channel 2D image as the input of a convolutional neural network which predicts the final market state of the given trading day.
The model provides accurate predictions on the S\&P index when compared with a random null model. 

The promising results observed in this challenging financial context - with hardly predictable data and a limited set of relevant features - constitute a solid basis for further applications of this method to other noisy sequential data characterising complex systems. 
The method has been shown to overcome limitations for noisy time series with low predictability and could outperform other methodologies for more predictable data, such as biological and medical data, e.g. ECG signals, or weather and climate data, e.g wind speed and temperature time series. 
\section{Materials and Methods}
\subsection{Mother wavelet}
\begin{figure}[htbp]
    \centering
    \includegraphics[width=7cm]{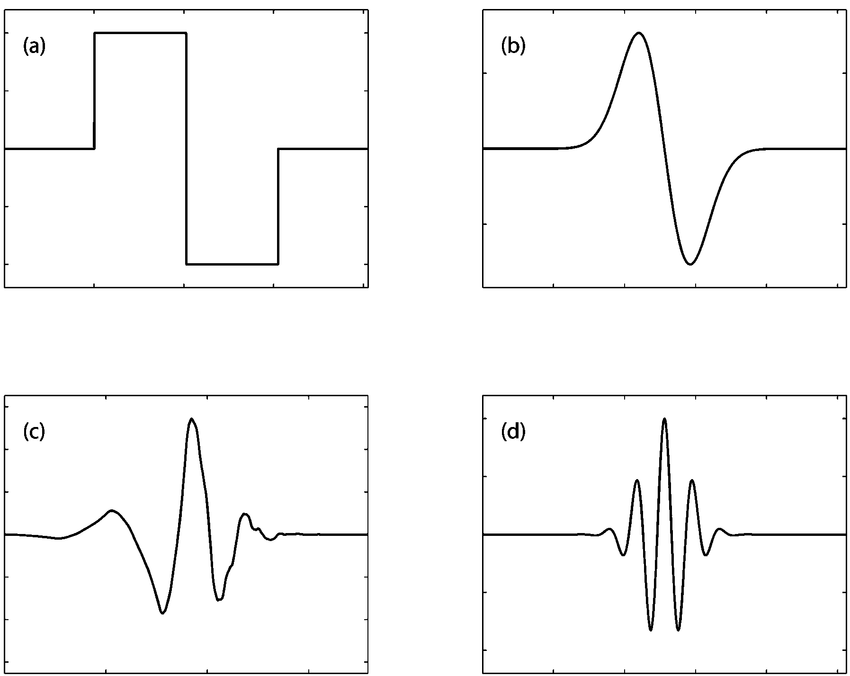}
    \caption{Four commonly used mother: wavelets (a) Haar wavelet, (b) Gaussian wavelet of order 1, (c) Daubechies wavelet of order 4, and (d) Morlet wavelet.\cite{baker2007quantitative}}
    \label{fig:Four Wavelets}
\end{figure}

Figure \ref{fig:Four Wavelets} shows the four mother wavelets. The Haar wavelet(a) and (c) Daubechies wavelet of order 4 are discrete wavelets. (b) Gaussian wavelet of order 1 and (d) Morlet wavelet are continuous wavelets. 
\subsection{Threshold Methods}
Hard Thresholding
\begin{equation}
\sigma_{\lambda}^{H}(w)=\left\{\begin{array}{l}
w,|w| \geq \lambda \\
0,|w|<\lambda
\end{array}\right.
\label{eq:hard thresholding}
\end{equation}
Soft Thresholding
\begin{equation}
\sigma_{\lambda}^{s}(w)=\left\{\begin{array}{c}
{[\operatorname{sgn}(w)(|w|-\lambda)],|w| \geq \lambda} \\
0,|w|<\lambda
\end{array}\right.
\label{eq:soft thresholding}
\end{equation}
The $w$ is the wavelet coefficient (detail coefficient). The $\lambda$ is the selected threshold.
\subsection{Rigrsure threshold}
Rigrsure threshold is an adaptive threshold selection using the principle of Stein's Unbiased Risk Estimate (SURE).\\ 

(1)Take the absolute value of the elements in the signal $\Psi[t]$, and then sort from small to large, square each element to get a new signal sequence f(k)\cite{valencia2016comparison}.
\begin{equation}
f(k)=(\operatorname{sort}(|\Psi|))^{2}, \quad(k=0,1, \ldots, N-1)
\end{equation}
(2)If the threshold is the square root of the element of the k-th element of $f(k)$, \begin{equation}
\lambda_{k}=\sqrt{f(k)}, \quad(k=0,1, \ldots, N-1)
\end{equation}
the risk generated by the threshold is
\begin{equation}
\operatorname{Rish}(k)=\left[N-2 k+\sum_{i=1}^{k} f(j)+(N-k) f(N-k)\right] / N
\end{equation}
(3)According to the obtained risk curve $Risk(k)$, take the $k_min$ corresponding to the minimum risk point lock, then the rigrsure threshold is
\begin{equation}
\lambda_{k}=\sqrt{f\left(k_{\min }\right)}
\end{equation}

\subsection{Return frequency distributions}
\begin{figure}[htbp]
    \centering
    \includegraphics[width=12cm]{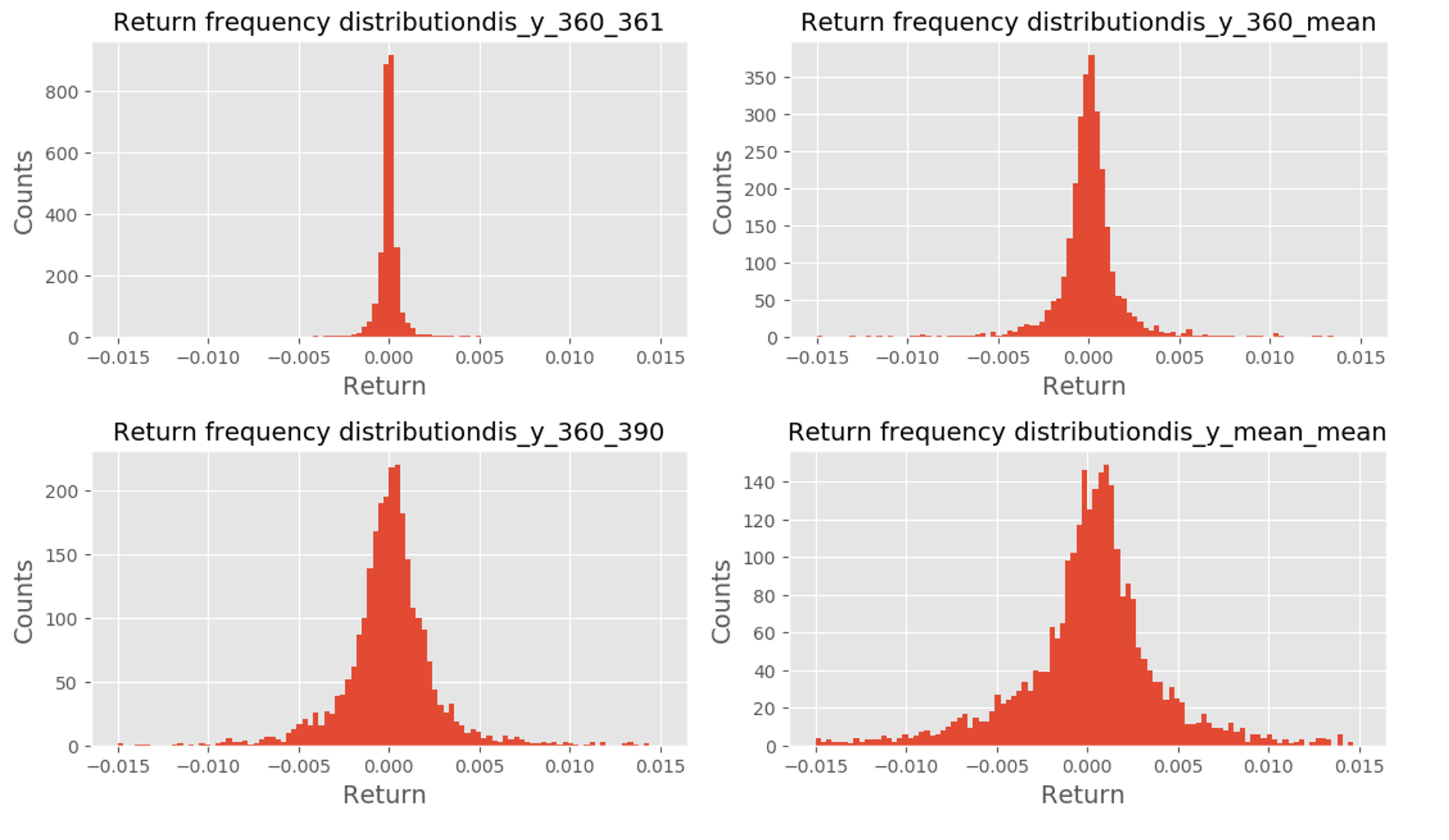}
    \caption{The return frequency distributions another four label designs.}
    \label{fig: other labels}
\end{figure}

Figure \ref{fig: other labels} shows the return distributions for four different label designs. $y_{360-361}$: the 360th minute price compare with the 361th minute price. $y_{360-mean}$: the 360th minute price compare the average price of the last 30 minutes. $y_{360-390}$: the 360th minute price compare with the 390th minutes market closing price. $y_{mean-mean}$: the average price from 1st minutes to 360th minutes compare with the average price of the last 30 minutes

\bibliographystyle{unsrt}  
\bibliography{references}  





\end{document}